\documentclass[letterpaper]{article}

\usepackage{natbib,alifeconf}  
\usepackage{url,hyperref,cleveref}
\usepackage{booktabs}
\usepackage{amssymb}

\title{ALIFE2024 template}

\title{Non-Platonic Autopoiesis of a Cellular Automaton Glider in Asymptotic Lenia}

\author{
    Q. Tyrell Davis$^{1}$,
    \mbox{}\\
    $^1$Independent Researcher\\
    Boulder, Colorado, United States \\
    qtd.science.wrought049@passmail.net
} 
\begin{document}

\maketitle

\section{Introduction}

\begin{figure}[t!]
    \centering
    \includegraphics[width=3.25in,angle=0]{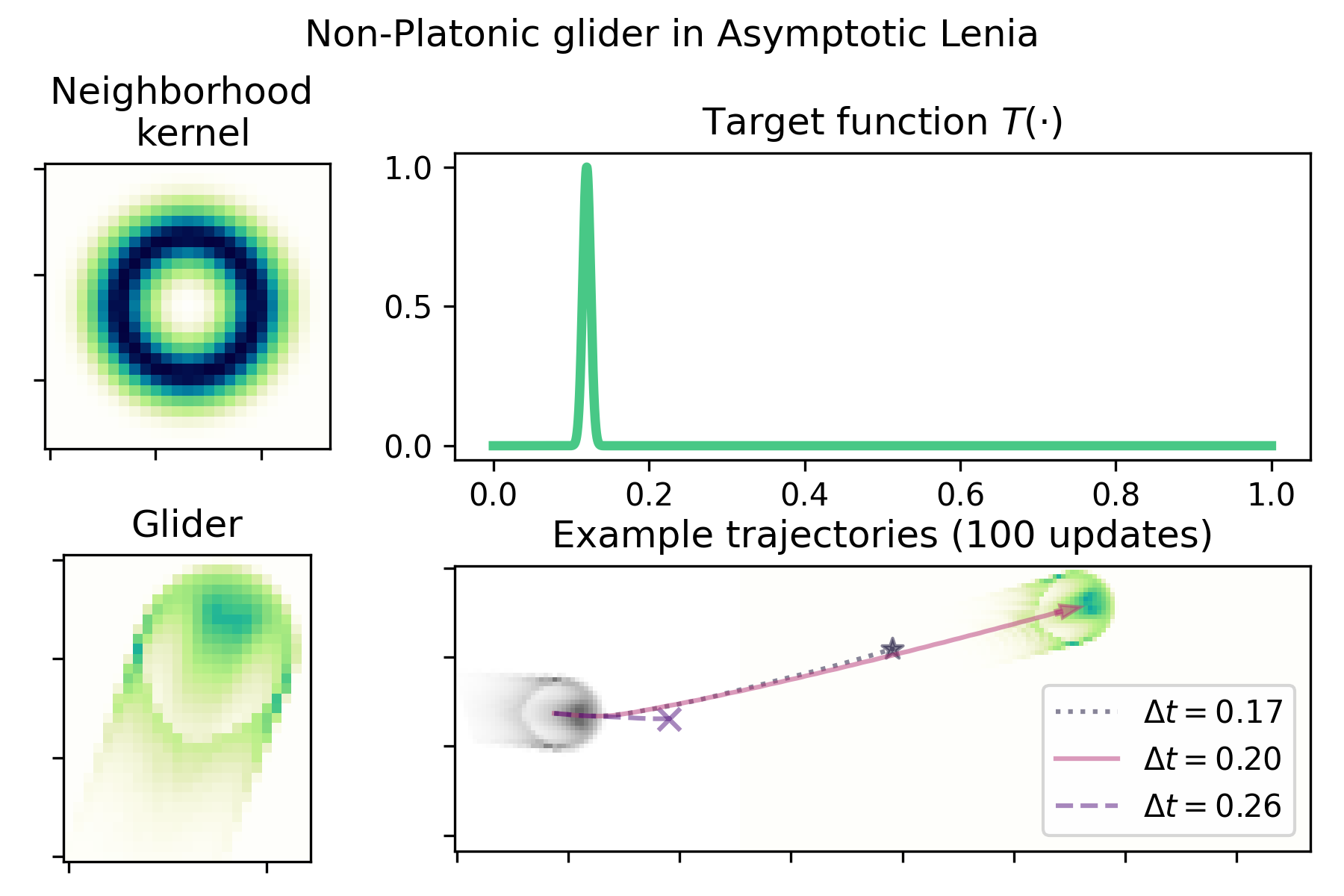}
    \caption{
        {\bfseries A non-Platonic glider in asymptotic Lenia.} With step size $\Delta t=0.2$, the glider persists for 100 update steps, but disappears earlier for step sizes $\Delta t=0.17$ or $0.26$. The rules supporting this glider are defined by a target function (Equation \ref{eqn:target})
        with $\mu_T = 0.12$ and $\sigma_T = 0.005$, and neighborhood kernel $K$ with $\mu_K = 0.5$ and $\sigma_K = 0.15$ and native radius $k_0 = 13$ pixels ($r=1.0$ at 13 pixels from kernel center). Code available at \href{https://github.com/riveSunder/fractal_persistence/}{https://github.com/riveSunder/fractal\_persistence}. Instructions for replicating results at commit 09437415... 
    }
    \label{fig:summary}
\end{figure}

Like Life, Lenia CA support a range of patterns that move, interact with their environment, and/or are modified by said interactions. These patterns maintain a cohesive, self-organizing morphology, {\itshape i.e.} they exemplify autopoiesis, the self-organization principle of a network of components and processes maintaining themselves \citep{varela1974}, and a framework for investigating the life-like properties of CA \citep{beer2004, beer2015}. 

Lenia dynamics are described by PDEs. This description lends itself to simulation with computers, as in classical physics, albeit incurring the `necessary evil' of discretization. Unlike simulating classical physics, where a trade-off of precision and accuracy in exchange for keeping time and computational costs feasibly low is made, a subset of patterns in Lenia and similar systems have been shown to lose their autopoietic competence when simulated with discretization parameters too fine \citep{davis2023, kojima2023,  davis2022b}. For these patterns in the context of their respective rules, simulation appears part and parcel to their ability to self-organize: they exhibit non-Platonic autopoiesis.


Lenia has been expanded with a number of variations on the original. In this work I examine a glider in asymptotic Lenia (abbreviated `ALenia' in this work) \citep{kawaguchi2021}. ALenia replaces the growth function with a target function: the update rule is then defined by the difference between the output of the target function and the current cell states. 

Standard Lenia uses a clipping function to constrain cell values to a range from 0 to 1, but ALenia's update scheme does away with the need. Recent work implementing ALenia as a reaction-diffusion system reported that non-Platonic behavior in standard Lenia may depend on the clipping function, and that ALenia gliders are likely not affected by non-Platonic dissolution \citep{kojima2023}. In this work I show an example of a glider in ALenia that depends on a certain simulation coarseness for autopoietic competence: when simulated with too fine spatial or temporal resolution the glider no longer maintains its morphology or dynamics. 

\section{Lenia and Asymptotic Lenia}
An instance of Lenia has states $A$ defined as cell (or pixel) values at coordinates $x = (x_h, x_w)$ that change over time $t$. The update rule has a form similar to the Euler method, adding the product of the result of differential equation $\frac{\partial A}{\partial t}$ and step size $\Delta t$:

\begin{equation}
    A(x,t+\Delta t) = \left[ A(x, t) + \Delta t \cdot \frac{\partial A}{\partial t}\right]_{0}^{1}
    \label{eqn:lenia}
\end{equation}

Unlike the Euler method, Equation \ref{eqn:lenia} constrains cell values $A(x,t)$ in the range $\mathcal{R}[0,1]$. Hard truncation, represented by $[\cdot]_0^1$, is typically used, but many pattern-rule pairs work equally well with a smooth squashing function. 

Lenia dynamics are defined by a differential equation, aka the growth function $G(\cdot)$.

\begin{equation}
    \frac{\partial A}{\partial t} = G(K \circledast A(x,t)) 
    \label{eqn:lenia_dadt}
\end{equation}

Where $K$ is a neighborhood kernel ({\itshape e.g.} top left in Figure \ref{fig:summary}) and $\circledast$ represents convolution. The growth function takes a similar form to that of a Gaussian, stretched to output values from -1 to 1: 

\begin{equation}
    G(x) = 2 \cdot e^{-\left(\frac{x - \mu_G}{\sigma_G}\right)^2} - 1
    \label{eqn:lenia_growth}
\end{equation}

Where $\mu_G$ and $\sigma_G$ set the peak and width of the curve and $x = K \circledast A(x,t)$ represents cell neighborhoods.

The neighborhood kernel $K(r) = e^{-\left(\frac{r - \mu_K}{\sigma_K}\right)^2}$ is also Gaussian-like, and acts on radial coordinates $r$. Neighborhood kernels are normalized to sum to 1.0.

\if{0}
    \begin{equation}
        K(r) = e^{-\left(\frac{x - \mu_K}{\sigma_K}\right)^2}
        \label{eqn:lenia_kernel}
    \end{equation}
\fi

ALenia replaces the growth function $G$ with a target function $T$. $\frac{\partial A}{\partial t}$ is then the difference of cell values $A(x,t)$ and target function applied to cell neighborhoods $K \circledast A(x,t)$

\begin{equation}
    \frac{\partial A}{\partial t} = T(K \circledast A(x,t)) - A(x,t)
    \label{eqn:alenia_dadt}
\end{equation}

ALenia's target function is unstretched, yielding values from 0 to 1: 

\begin{equation}
    T(x) = e^{-\left(\frac{x - \mu_T}{\sigma_T}\right)^2}
    \label{eqn:target}
\end{equation}

 As a result of the asymptotic target function, ALenia obviates the need to truncate cell values.


\section{Autopoiesis and Discretization}

\begin{figure}[h!]
    \centering
    \includegraphics[width=3.125in,angle=0]{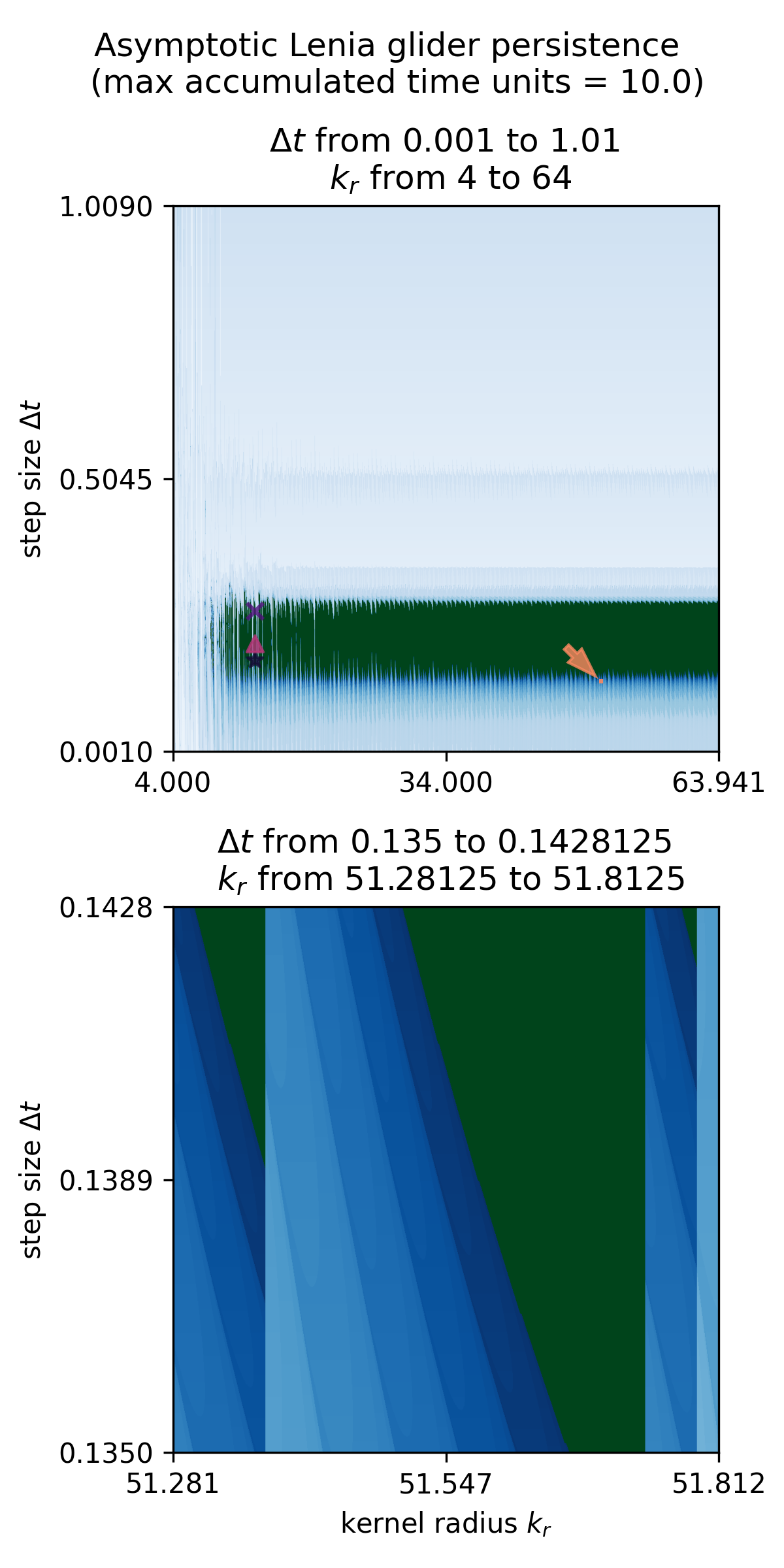}
    \includegraphics[width=3.125in,angle=0]{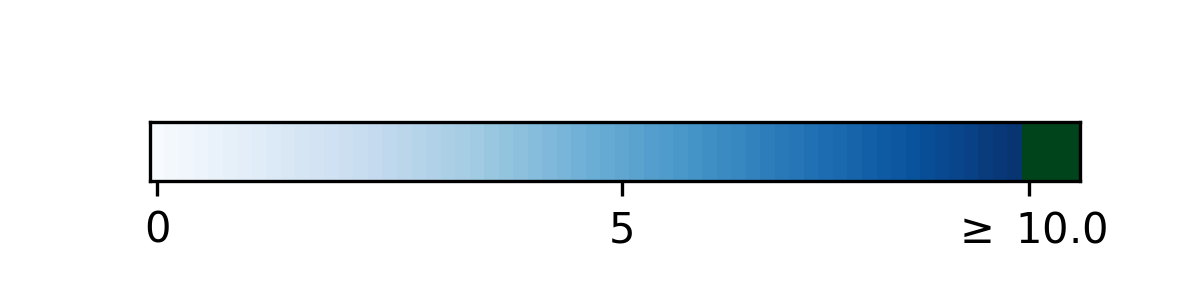}
    \caption{
        {\bfseries ALenia glider persistence with respect to $\Delta t$ and $k_r$ shows non-Platonic loss of autopoiesis}. Persistence in accumulated time units is indicated by dark colors, with a transition from blue to green at the arbitrary threshold of 10 time units, the maximum simulation time.  
        `X', triangle, and star symbols mark $\Delta t = 0.26$, $0.20$, and $0.17$ from Figure 1, respectively. The lower persistence map is a zoom to the region in the small orange rectangle indicated by the arrow, showing retained details of apparent fractal structure.  
    }
    \label{fig:nonplatonic}
\end{figure}

I simulated the glider in Figure \ref{fig:summary} across a range of spatial and temporal resolutions defined by $\Delta t$ and $k_r$, respectively. 

I modified temporal resolution $\Delta t$ by simply using a different value for the step size \texttt{dt}. In pseudocode: \texttt{A = A + dt * (T(conv(A,K)) - A)}. To smoothly vary spatial resolution, I change the size of the neighborhood kernel $K(r)$ by normalizing to kernel radius $kr$ (in pixels): \texttt{xx, yy = meshgrid(x, y); r = sqrt(xx**2 + yy**2) / kr}. I used scikit-image \citep{van2014} to rescale patterns\footnote{\texttt{skimage.transform.rescale} with \texttt{scale\_factor = kr/k0} and \texttt{order=5} spline interpolation, and  \texttt{anti\_aliasing=True} when downscaling.} and define glider stability by homeostasis of glider `mass', {\itshape i.e.} the sum of cell values. I chose normalized (relative to starting) cell-sum thresholds of 0.9 to 1.3 based on empirical observations of typical variation for a stable glider.

Results are shown in Figure \ref{fig:nonplatonic}. Gain of stability when moving from fine to coarse simulation parameters ($\Delta t$, $k_r$) is consistent with non-Platonic autopoiesis, as is a loss of stability when increasing $k_r$ and/or decreasing $\Delta t$.

\footnotesize

\clearpage

\section{Appendix: The Eidolic Principle: Exploring the Fractal Simuliverse in a Continuous Cellular Automaton}
%
%

\section{Introduction}

\begin{figure}[t!]
    \centering
    \includegraphics[width=1.5in,angle=0]{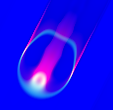}
    \caption{
        {\bfseries Adorbium}, {\small a CA pattern that resides in an adaptive moment estimation CA. Red channel is grid states, green is the first moment $m$ and, and blue is the second moment $v$. Adorbium CA parameters are $\beta_1=0.8$, $\beta_2=0.99$, $\epsilon=10^{-8}$, $\mu_G=0.167$, $\sigma_G=0.013$, $\mu_K=0.5$, $\sigma_K=0.15$, and kernel radius $k_r=13$ pixels. This document accompanies a video exploration of this glider's autopoietic simuliverse available at \href{https://youtu.be/pH1x-6FzmTo}{https://youtu.be/pH1x-6FzmTo}. Code is available at \href{https://github.com/riveSunder/fractal_persistence/}{https://github.com/riveSunder/fractal\_persistence/}}
    }
    \label{fig:adorbium}
\end{figure}

The Anthropic Principle posits that only a universe in which life and intelligent observers emerge can be consciously observed \citep{dicke1957, carter1974}. The principle reminds us that while it may seem that the universe and its physics are tailored to us, our perspective is biased. 


\begin{figure}[hb!]
    \centering
    \includegraphics[width=3.125in,angle=0]{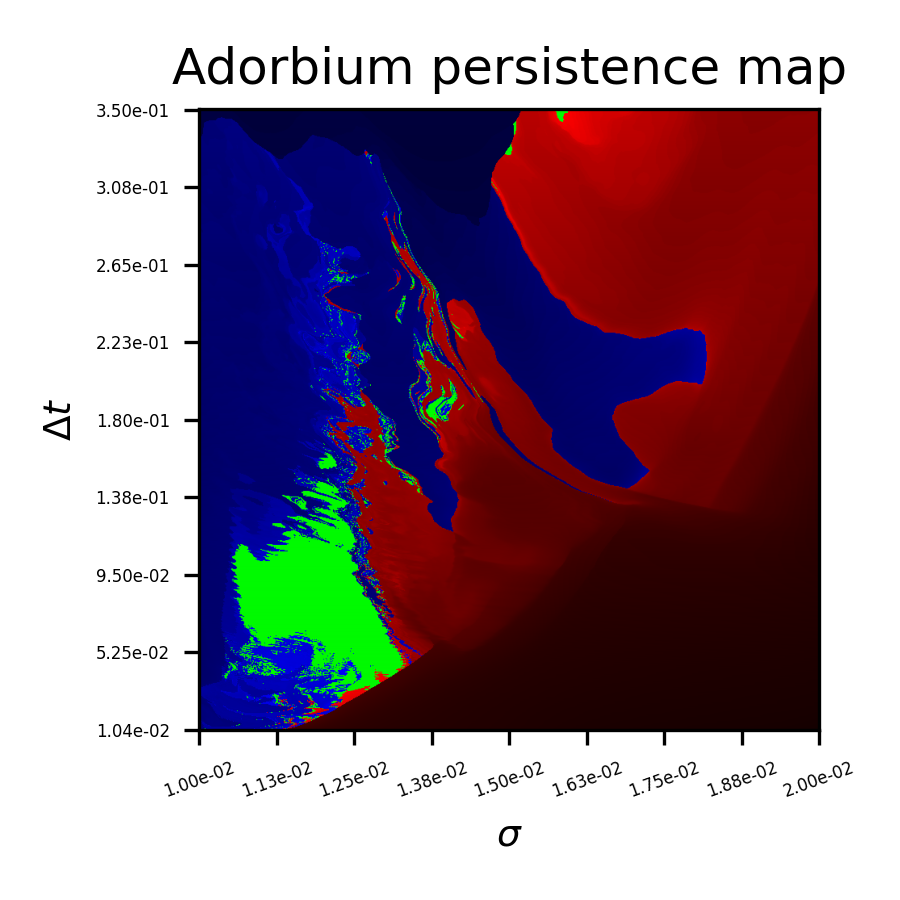}
    \caption{
        {\bfseries Persistence map} {\small Red channel indicates space-filling and blue channel vanishing loss of persistence, while the green channel indicates persistence  up to an arbitrary max simulation time of 16.0 simulation time units. Luminance is proportional to accumulated time units.}
    }
    \label{fig:persistence}
\end{figure}

The Eidolic Principle similarly notes the the specificity of physical parameters supporting self-organizing patterns (such as gliders) in simulations of continuous cellular automata (CA). Only one or a few fundamental self-organizing patterns are typically found in a given point in the parameter space of a continuous CA.

Self-organizing patterns in continuous CA depend on implementation details as well as physical parameters, and can lose the ability to self-organize at both too coarse {\itshape and} too fine simulation settings \citep{davis2022b, kojima2023, davis2023}. 

This work explores the persistence map (Figure \ref{fig:persistence}) of a CA pattern, the boundaries of which retain intricate fractal detail across several orders of magnitude. The pattern resides in a CA framework based on adaptive moment estimation optimization (adam)\citep{kingma2014}. The pattern (Figure \ref{fig:adorbium} is reminiscent of, and shares a neighborhood kernel with, {\itshape Orbium} gliders from the continuous CA framework Lenia \citep{chan2019}. 

The adam CA framework updates grid states $A(x,t)$ by first calculating the first moment $m_{t+\Delta t} = \beta_1 m_t + (1-\beta_1) G(K_n \circledast A_t)$, second moment $v_{t+\Delta t} = \beta_2 v_t + (1-\beta_2) G(K_n \circledast A_t)^2$. $m$ and $v$ are then used to calculate the update function $ \frac{\partial A}{\partial t} = \frac{m}{\sqrt{v}+\epsilon}$, which modifies grid states $A_t$ proportional to step size $\Delta t$.

\begin{equation}
    A_{t+\Delta t} = A_t + \Delta t \frac{m}{\sqrt{v}+\epsilon}
    \label{eqn:adam_update}
\end{equation}

\subsection{Why do CA Persistence Maps Exhibit Fractal Detail Across Scale?}

Cellular automata that support `interesting' dynamics\footnote{While the notion of interesting is ambiguous to define, in CA we often think of self-organization and complex interactions, as exhibited quintessentially by gliders, as worth investigating.)} are characterized by variation between chaos and order over space and time. A hallmark of chaotic behavior is sensitivity to arbitrarily small differences in initial conditions. This appears to be the case where the boundary between CA glider persistence and instability remains rough down to the limits of floating point precision. Fractal patterns have likewise been shown for neural network trainability with respect to hyperparameters \cite{sohldickstein2024} (a significant influence on the work presented here) and fractal structure is likely in other complex systems with outputs that are {\itshape a priori} undecidable.

\newpage

\footnotesize
\bibliographystyle{apalike}
\bibliography{references} 

\begin{thebibliography}{}

\bibitem[Beer, 2004]{beer2004}
Beer, R.~D. (2004).
\newblock Autopoiesis and cognition in the game of life.
\newblock {\em Artificial Life}, 10(3):309--326.

\bibitem[Beer, 2015]{beer2015}
Beer, R.~D. (2015).
\newblock Characterizing autopoiesis in the game of life.
\newblock {\em Artificial life}, 21(1):1--19.

\bibitem[Carter, 1974]{carter1974}
Carter, B. (1974).
\newblock Large number coincidences and the anthropic principle in cosmology.
\newblock In {\em Symposium-international astronomical union}, volume~63, pages 291--298. Cambridge University Press.

\bibitem[Chan, 2019]{chan2019}
Chan, B. W.-C. (2019).
\newblock {Lenia} - biology of artificial life.
\newblock {\em Complex Systems}, 28:251--286.

\bibitem[Davis, 2023]{davis2023}
Davis, Q.~T. (2023).
\newblock Discretization-dependent dissolution of gliders in (dis)continuous systems: Non-platonic self-organization in complex systems.
\newblock {\em Innovations in Machine Intelligence (IMI)}, 3:1--23.

\bibitem[Davis and Bongard, 2022]{davis2022b}
Davis, Q.~T. and Bongard, J. (2022).
\newblock Step size is a consequential parameter in continuous cellular automata.
\newblock {\em ALIFE 2022: The 2022 Conference on Artificial Life}, 43.

\bibitem[Dicke, 1957]{dicke1957}
Dicke, R.~H. (1957).
\newblock Gravitation without a principle of equivalence.
\newblock {\em Reviews of Modern Physics}, 29(3):363.

\bibitem[Kawaguchi et~al., 2021]{kawaguchi2021}
Kawaguchi, T., Suzuki, R., Arita, T., and Chan, B. (2021).
\newblock Introducing asymptotics to the state-updating rule in {Lenia}.
\newblock In {\em ALIFE 2021: The 2021 Conference on Artificial Life}. MIT Press.

\bibitem[Kingma and Ba, 2014]{kingma2014}
Kingma, D.~P. and Ba, J. (2014).
\newblock Adam: A method for stochastic optimization.
\newblock {\em CoRR}, abs/1412.6980.

\bibitem[Kojima and Ikegami, 2023]{kojima2023}
Kojima, H. and Ikegami, T. (2023).
\newblock Implementation of {Lenia} as a reaction-diffusion system.
\newblock In {\em ALIFE 2023: Ghost in the Machine: Proceedings of the 2023 Artificial Life Conference}. MIT Press.

\bibitem[Sohl-Dickstein, 2024]{sohldickstein2024}
Sohl-Dickstein, J.~N. (2024).
\newblock The boundary of neural network trainability is fractal.
\newblock {\em ArXiv}, abs/2402.06184.

\bibitem[Van~der Walt et~al., 2014]{van2014}
Van~der Walt, S., Sch{\"o}nberger, J.~L., Nunez-Iglesias, J., Boulogne, F., Warner, J.~D., Yager, N., Gouillart, E., and Yu, T. (2014).
\newblock scikit-image: image processing in python.
\newblock {\em PeerJ}, 2:e453.

\bibitem[Varela et~al., 1974]{varela1974}
Varela, F.~G., Maturana, H.~R., and Uribe, R. (1974).
\newblock Autopoiesis: The organization of living systems, its characterization and a model.
\newblock {\em Biosystems}, 5(4):187--196.

\end{thebibliography}

\end{document}